# Ontological Categorizations and Selection Biases in Cosmology: the case of Extra Galactic Objects

P. Valore[1], M.G. Dainotti[2,3,4], O. Kopczyński[3]

**Abstract:** One of the innovative approaches in contemporary philosophical ontology consists in the assumption of a plurality of ontologies based on different metaphysical presuppositions. Such presuppositions involve, among others, the identification of relevant properties for the objects of our domain as a guiding principle in uncovering what it is to be considered intrinsic and what could be the mere effect of selection preferences based on objective or subjective criteria. A remarkable example of the application of a background metaphysical theory in astrophysics is the problem of selection biases in detecting cosmological objects, such as supernovae, galaxies and gamma-ray bursts. We will show that it is valuable to be aware of the importance of uncovering this type of background theory to better understand selection effects and to promote a novel approach in scientific research.

1. Introduction

The task and scope of "ontology" have been negotiated many times. In particular, the very meaning of "ontology" can vary to a significant extent: the mere specification of a taxonomy (Koepsell 1999); the formal conceptualization, i.e. a description (as in the formal outline of a program) of the concepts and relations allowed within a language (Gruber 1993); the philosophical discipline that is (partly) overlapping with metaphysics (Hofweber 2020, Varzi 2011). Most of the applied ontology takes its task to be a description of domain-specific ontologies. Applications in this sense can be found in several fields: computer science and philosophy of computing and information (Guarino-Giaretta 1995, Guarino 1998, Smith 1998, Smith 2003), bio-medical disciplines (Gangemi *et al.* 2003, Valore 2017a), taxonomical sciences, theology, social sciences, and so on; sometimes with a significant contribution of mereology (Varzi-Cotnoir in press), the study of the relationship between whole and parts (e.g. in biomedical sciences: Bittner, 2004; in topology: Smith, 2006 ). Furthermore, there is more than one correct paradigm, more than one way to build ontological taxonomies and different ways to ground ontology in metaphysics (see Valore 2016, Garcia 1999). A comprehensive discussion of alternative systems of ontological categories and of the available foundations of ontology is out of the scope of our paper. Moreover, the distinction of various definitions and tasks of "ontology" doesn't exclude that traditional philosophical ontology may be continuous with applied ontology in the sense of a description of domain-specific ontologies (Smith 2003, Guarino-Carrara-Giaretta 1994). Nonetheless, to avoid terminological confusion and to clear the field from assumptions that we do not mean to make, here we decided to stick to the philosophical meaning defended by Quine 1980 and Van Inwagen 2001, and to adopt the so called "standard paradigm" (Fine 2009) in the study of *philosophical ontology*, i.e. the approach that considers the specification of an ontology as the specification of which entities (or, rather, which *classes*, *categories* of entities) are required by the truth of the sentences of a scientific theory once they are translated in quantification theory.

The development of such a systematic investigation of a domain of entities and its categorial taxonomies has given birth to an "ontological turn" which, in the last few decades, has grown to encompass all branches of philosophical research. This expansion of ontology has been evaluated (e.g. Azzouni 1998, 2010 and Fine 2009, 2001 extensively discuss whether the quantificational form is a satisfactory ontology, in the sense that it satisfies all the desiderata of traditional ontology) and challenged (Hirsch 2011; Price 2009; Thomasson 2015; Schaffer 2009; Tahko 2012; Tahko 2015; Correia & Schnieder 2012; Schnieder, Steinberg & Hoeltje 2013; Novotný & Novák 2014). However, the "metaphysical" background of the quantification model remains generally unquestioned (see Valore 2017b). A recent and promising innovative approach seems to be an improved accounting of the (sometimes hidden) metaphysical, background presuppositions of our ontological categorizations. The final goal is to contribute to contemporary research in ontology and metaphysics from a pioneering point of view, one that finally discloses the background "metaphysical" assumptions and conceptual "bias" of our ontology. Such presuppositions involve, among others, the identification of relevant properties for the objects of our domain as a guiding principle in uncovering what it is to be considered intrinsic and what could be the mere effect of selection preferences in building the correct classes of objects.

---

[1] University of Milan, Department of Philosophy, Milan, Italy; paolo.valore@unimi.it
[2] Stanford University, Kavli Institute for Particle Physics and Cosmology (KIPAC), Stanford, USA, mdainott@stanford.edu
[3] Astronomical Observatory of Jagiellonian University, Krakow, Poland; oskar.k@poczta.onet.pl
[4] Space Science Institute, Space Science Institute - 4765 Walnut St, Suite B, Boulder, CO 80301

In order to show how neglecting the metaphysical background theory that validates these preferences may impact scientific research and to advocate for cross-fertilization among astrophysical studies and philosophical conceptual analysis of ontological categorization, we will consider the notion of "intrinsic properties" in astrophysics. Here, it is assumed that i) the intrinsic properties are those that exist independently from the selection effects that can bias them and ii) if we believe we have properties that are affected by biases, we need to find a way so that we can finally be able to reveal the underneath intrinsic properties.

A notable and clear illustration of the application of a background metaphysical theory in astrophysics is the problem of selection biases in detecting cosmological objects, such as supernovae, galaxies and gamma-ray bursts. In this paper, we will examine the impact of categorization and conceptual assumptions in parameters used to describe extra galactic objects. We will address the control of selection biases when handling the problem of the evolution of the observables in a multi-dimensional space, focusing on the case of Gamma Ray Bursts (GRBs), which are among the farthest events ever observed in the Universe.

## 2. Ontological categorization and metaphysical assumptions

Ontology addresses the question of what exists. Many contemporary ontologists take this to be a project of specifying the categories of objects that we should assume, given a certain theory and its associated domain of individuals. This way, philosophical ontology displays the link between our true theories and the general kinds of entities that we can assume in our universe. The general categories of our ontology are to be found in the upper level of a flux diagram that represents the generality levels and the subordination relationship between nodes (kinds of things) and between nodes and leaves (individuals). Drawing this tree can be seen as essentially mirroring the instantiation of categories, *via* kinds of lesser and lesser generality, by individuals. Examples of ontological flux diagrams can be found in Chisholm 1996, Hoffman & Rosenkrantz 1994, the BFO project in Smith 1998, the DOLCE project in Gangemi *et al.* 2003 and many others (for a presentation of some systems, cf. Valore 2016). We will *not* assume here any specific categorial system, nor will we offer any project of domain-specific ontology, based on a particular ontological flux diagram. What we want to focus on here is that, when we try to justify the link between individuals and categories within *any* ontological tree, we assume the notion of exemplification based on property sharing. In other terms, two or more individuals exemplify the same general kind because they are "similar" under a certain respect; and the notions of the individuals being "similar" or the individuals having "something in common" can be reduced to manageable notions in terms of set theory. Assuming the set of individuals $I$ associated to a given theory and a set of predicates $P$, we are requested, for any member of the first set, to pair the appropriate predicate or the negation of the predicate. This way, given the two sets $I = \{a, b, …, n\}$ and $P = \{P_1, P_2, …, P_m\}$, we can identify the subset $S_i$ of I selected thanks to $P_j$.

There are several oversimplifications in this picture, known in literature (e.g. Putnam 1999, Valore 2016, 2017a, 2018). In order to reduce the similarity among entities to set theory notions, we are assuming:
i) to be always possible, for any $I_k$, to determine whether $P_j$ holds, and
ii) properties do not come in degree, i.e. that properties are either enjoyed or not enjoyed, assuming a binary logic. However, in science we are aware that some simplifications may affect the use of classes for categorizing objects; for example, the case of the classification of short and long Gamma-Ray Bursts and of the several classes of Supernovae which will be discussed in the following sections.

Furthermore, there seem to be at least two types of similarity at play when we try a categorization of a population of individuals assuming the notion of "having something in common":

a) we can build kinds assuming an object-to-object resemblance and

b) we can link individuals to kinds, thanks to an object-to-kind resemblance. The two types of similarities work jointly: we focus on resemblances among individuals to build general cases, which result from selective acts of grouping for the sake of our representation, or conceptualization and, by relying on the general cases thus built, we are able to associate other individuals with kinds to which they are similar. This is particularly relevant in the framework of the similarities among Gamma-Ray Bursts, supernovae and galaxies to each other their own kinds.

However, even if we decide to grant these oversimplifications for the sake of a more straightforward (albeit less accurate) analysis, there seems to be a more serious complication on our way.

An attempt to improve this strategy can be developed by means of a translation such as: $\{[I_1$ is more *similar* to $I_2$ than $I_3] =_{Def} [I_1$ and $I_2$ are jointly members to more sets than $I_1$ and $I_3]\}$. This model is known not to work (it was, for instance, dismissed already in Quine 1969). In set theory we admit that every collection of things forms a set. Thus, since sets are generated by the exhaustion of combinations, the number of sets to which any two elements jointly belong isn't determined by those elements' properties (i.e. their similarity), but by the total quantity of elements. In some cases, when we do not know what are all the relationships that should hold among objects, a blind search (considering all possible combinations) can unveil relations which were not considered before. However, in general, if our purpose is to give a representation of the relationships between individuals and kinds *based on how the things are* (i.e. their properties), clearly, we shouldn't consider all the combinations in sets that are possible given a certain population of individuals.

A selection of meaningful combinations would allow us to reduce the number of sets of things we need to consider in order to identify the proper kinds of things and to build the correct ontological tree, blocking the improper categorizations. The very notions of categories and kinds of things suggest that the individuals we group under certain respects exhibit the same sortal property since they share something like a *nature*, a *disposition*, or an *essence* (however defined). In order to do the job, we need the notion of *intrinsic* or *relevant* property. For example, to compare extragalactic objects, we need to find commonalities or differences between their intrinsic properties. The problem being, at this point, that accordingly to different theoretical assumptions, we may get different *intrinsic* or *relevant* properties. Simply stated, we cannot make sense of the notion of "relevant" without specifying what we assume it has to be relevant for. In addition, we may want to consider that properties change through time and our notion of persistence needs to be grounded in some way (and the philosophical literature has plenty of thought experiments and challenging questions about the notion of chronological identity notwithstanding changes in properties).

Consequently, there are many alternative ways of classifying things in classes according to their shared properties, producing different patterns of classification in categories guided by different theoretical assumptions about the correct properties and the correct philosophical conception about persistence through changes of properties. If we are not able to specify the strategy of the intended relevant properties and our preferences in advance, the empirical data themselves do not fully help us spot the perspicuous kinds. These theoretical assumptions offer the framework of the metaphysical background of any theory that assumes the notion of *intrinsic property* and *pertinent* and *meaningful category*.

3. **Categorization vs physical reality: the supernovae example**

Until the mid-1980s, we assumed a categorization of supernovae in two basic kinds: Types I and Types II, depending on the presence (Type II) or lack (Type I) of hydrogen lines in their observed spectra. The light curves (the brightness of the source as it evolves with time) of Type Is were all similar, and the common progenitor scenario for Type Is was assumed: a white dwarf companion of a red giant star accretes matter from the latter, until it reaches 1.4 solar mass limit. This causes a carbon thermonuclear detonation, which results in a supernova explosion. Type IIs were instead considered as explosions of evolved large mass stars due to the gravitational collapse of iron core (core-collapse supernovae). Yet in the late-1980s subtypes of Type I were distinguished, based on the more detailed spectral analysis: Type Ia, Ib and Ic. (Gal-Yam 2017) While Ias were considered the white dwarf explosions, as described above, Types Ib and Ic were considered a specific core-collapsed stars without a hydrogen envelope. Despite similar properties they turned out to be *entirely* different events, when physical scenarios were considered.

Also, the classical scenario for Type Ia supernovae were challenged after a while. It turned out that most likely the binary white dwarf-red giant systems cannot be progenitors in case of the historic Type Ia supernovae in neither our Galaxy nor Magellanic Clouds due to the lack of ejected red giant companion stars (Ruiz-Lapuente 2018). Today double degenerate progenitors (a binary white dwarf system, causing a supernova explosion when merging) is considered more likely in the case of most SN Ia. Yet the details are still unknown, and there are several scenarios that must be considered (Maguire 2017, Livio & Mazzali 2018).

In addition, the Type Ia SNe were commonly used as standard candles, due to their relative constant brightness, when corrected for some systematic effects, like the evolution of the "stretch" parameter of the lightcurve.

### 4. Cosmology and model-derived properties for high redshift objects

The scientific examination of most physical objects is based on their "direct" observation, and their physical properties, which are defined in various ways. Those physical properties are usually considered as observables, the quantities and parameters we can "observe" with the aid of our instruments. Their definition is often based on the technical capabilities of our instruments, what they can show us, and how we interpret their read-outs. The read-outs may be, for example, brightness in different filters (usually expressed in traditional astronomical magnitude system), fluxes in different wavelengths, or spectral features (e.g. spectral indices or so-called equivalent widths of either emission or absorption lines). The physical interpretation of these readouts is dependent on the physical theory behind the instrument operations and the assumption that our instruments (like a CCD camera in telescope focus) work correctly. However, due to selection biases, we shouldn't infer a direct correspondence between observables and the ontological status of objects.

The same telescopes and filters are used for both the local Universe and cosmological-distance objects (like galaxies or supernovae). This makes their comparison tricky, as we need to consider that properties of extragalactic objects evolve with redshift. For example, the optical brightness of an object is usually expressed as magnitude in an appropriate filter. There are several systems of filters (with filters usually denoted by a letter, most popular are either UBVRI, or ugriz systems), see (Bessel 2005) for a comprehensive review. The filter boundaries and sensitivities as a function of wavelengths are defined in the local reference frame. However, due to the cosmological evolution, we need to obtain the "true" magnitudes in a given filter, and thus to correct for the so-called k-correction. But computing the k-correction is not that simple. We need to restore the proper magnitudes in each band, and we cannot perform this relying only on observables. Instead we must reconstruct each magnitude using Spectral Energy Distribution (SED) fitting, using an appropriate model for galaxy spectra. This model takes into account the sources of light in a different galaxy. The reconstructed properties are only as reliable as the underlying model, which is based on several assumptions and parameters. In particular, there are many assumptions concerning Star Formation History (SFH -describing the rate at which new stars are born as a function of time, often assumed as exponential decline) of a given galaxy, which form has to be assumed a priori, with the exact parameter values fitted.

This reconstruction depends not only on the physical description of given galaxy, but also on the cosmological model adopted. In almost every case the standard Friedmann-Lemaitre-Robertson-Walker (FLRW see e.g. Schneider 2006) model is adopted. It contains the input parameters $H_0$, $\Omega_k$, $\Omega_m$, $\Omega_\lambda$, representing today's Hubble constant, curvature, matter density, and dark energy. The different values of these parameters result in different values for calculated physical properties of cosmological objects

(especially those correlated with luminosity and physical size). Particularly crucial is the Hubble constant, as distances to cosmological objects and the age of the Universe at different redshifts are proportional to its value. The Hubble methods may be estimated by various methods either utilizing various objects and properties in the relatively low redshift Universe (especially comparing the so called luminosity distance with redshift for objects with known brightness, the so called standard candles), or via the analysis of Cosmic Microwave Background (CMB a thermal radiation with a temperature equivalent of about 2.7 K, coming from all direction of space) properties. While it once seemed that the results obtained by different methods converge within error intervals, in recent years it appeared that the values obtained from the CMB analysis from the Planck measurement (Planck 2018) and measurement of SNe Ia significantly diverge (Riess et al. 2016). The reason of this divergence is unknown. It may result from either the systematic errors we are not yet aware and do not take into account or that the FLRW model is only the first approximation of the cosmology of the Universe.

5. **Conceptual aspects of parameters used to describe galaxies**

In case of galaxies, some fundamental parameters describing their physical properties are the Star Formation Rate (SFR, measured in solar masses per year) and Stellar Mass (measured in solar masses). Their interpretation is such that SFR should correspond to the rate (the number per unit of time) in which new stars of given mass are born, while Stellar Mass should correspond to the total mass of all the stars in a particular galaxy. Yet, the direct measurement of SFR and Stellar Mass is extremely challenging. The SFR and Stellar Mass are derived indirectly, either via Spectral Energy Distribution (SED, describing the distribution of energy emitted at different wavelengths of electromagnetic spectrum), modeling (see e.g. Gnedin e.t. al 2016 for reference), or some correlations with some observables (UV light, emission lines, IR, radio measurements etc.). All approaches depend on some assumptions regarding the physics of either the stars, or the stars and interstellar matter in the galaxies. Likewise, the stellar mass is also derived in a similar way. The assumptions on both SFR and Stellar Mass determination are built on analogies with the local Universe. Several key input parameters connected with both SFR and Stellar Mass e.g. Initial Mass Function (IMF, describing the distribution of stars of different masses at the moment of their birth), or stellar evolution tracks (describing how the SED and brightness of a particular star changes over time, from the birth till death) are derived from the research of stellar populations in our Galaxy and a few neighboring galaxies. The different IMF or SFH assumed results in different SFR and Stellar Mass. Nevertheless, it is possible to categorize the galaxies with regards to these derived parameters. The assumptions behind them, like the SFH scenarios, can be challenged, although they are convenient and result in satisfactory fitting to the data (see e.g. Lee et al. 2010). On the next level, one can analyze those quantities further, for example by checking the distribution of Stellar Masses and SFRs, and build further hypotheses based on these quantities.

6. **Why Extra Galactic Objects**

GRBs are the most explosive phenomena in the Universe after the Big Bang. In their short-lived phase, which usually lasts from a few seconds to minutes, they emit the same amount of energy as the Sun releases over its entire lifetime. Thus, they are detected out to gigantic distances (redshift, denoted with *z*), some of them as far back as when the first stars formed. Since they are among the farthest astrophysical objects visible across the Universe, their huge luminosities and their association with massive stars or binary merger systems make GRBs unique and powerful probes of stellar evolution (how individual stars evolved) and star formation (how star are formed) beyond the epoch of reionization, which is the era in which the matter in the universe is re-ionized.

Despite being observed for more than 50 years, GRBs are still mysterious phenomena regarding their physical mechanisms. So far there is not a unique picture able to pinpoint a theoretical model that can fully explain the general emission mechanism of GRBs. There is, in fact, no shortage of proposed origins: (hypotheses about their nature) e.g., the explosions of extremely massive stars, fusion of two neutron stars, or the spin down of magnetized massive stars (the so-called magnetars), see right panel of Fig. 2. The different origin relates to a different phenomenology (in the sense that different phenomenology may suggest different origin, but it is not always the case). Short GRBs, which spike and decay in less than 2 seconds, are possibly produced by the fusion of two neutron stars or a neutron star and a black hole. Long GRBs, which after the initial spike fade over hours or even weeks, are possibly associated with the collapse of massive stars. However, several sub-categories of Long and Short GRBs have been identified, which may arise from different progenitors or progenitors with different properties. These GRBs have different features in some of their properties. They are an intermediate class of GRBs called the short with extended emission that show mixed properties between short and long: the spectrum (the X-ray Flashes), their duration (the ultra-long GRBs have a duration of more than 1000 seconds) and a clear association with Supernovae (GRB-SNe). There has been a debate in the literature (Fynbo et al. 2006, Della Valle et al. 2006) whether or not there is a third class of GRBs which do not have the SNe associated to long GRBs even though GRBs are observed nearby.

The classical ontological problem of identity and persistency through change of properties appears once we consider the change of the properties of GRBs when calculating properties from an observer frame or from the rest frame. For instance, for the case of some long GRBs or short GRBs with extended emission, if they are observed at large distances in reality in their rest-frame (reference of the GRB at whatever distance the GRB is) they become short GRBs. Thus, it means that it is very challenging to identify a set of properties, $P_j$, that pertain to definite classes, since the classes themselves may change the sortal property they assign to their members. From an ontological point of view, we have either the option of changing the relevant properties for something to be a member of a certain type *or* to decide that the individual is no longer a member of that class. In other words, we can decide to revise the relevant properties that define the class or the assigned type identity of a certain individual: the choice is between the adjustment of property of the class or the establishment of a new class with new properties (for example the above mentioned class of long GRBs not associated with SNe). This clearly gets more and more complicated when we consider that properties change through time and the problem of persistence applied to ontological categorization and the selection of the intended relevant properties seems to become one of the most puzzling research topics of contemporary ontology.

In astrophysics we address the evolution in time by using a functional form that relates the property of the object with redshift, z, (which is also a measure of time, namely a large redshift corresponds to an earlier epoch of the universe). For example we can use a function $g(z)=(1+z)^k$ where the index k defines the evolution of any relevant property that can depend on the redshift pertinent to that object (Dainotti et al. 2013a, 2017b, 2015b; Dainotti & Del Vecchio 2017) or more complex functions.

We compute the evolution through statistical methods, the Efron & Petrosian method (Efron & Petrosian 1992) which allows for a way to remove instrumental selection biases and to overcome the problem of redshift evolution. The term selection effect or bias is referred to the distortion of statistical correlations, resulting from the method of collecting samples. Gathering data from a certain satellite with a certain flux limit prevents us from seeing a truly representative sample of events.

The Efron & Petrosian method is important also in determining the intrinsic properties of GRBs.

Indeed, there is a common motto in the high energy astrophysics community which is the following: "if you have seen one GRB you have seen just one" confounding our efforts to find commonalities among them.

Besides the characterization based on their duration, spectral properties and association with SNe, from a phenomenological point of view, a GRB emission at high photon energies (gamma-rays and hard X-rays) is called prompt, while the long lasting multi-wavelength emission, (X-ray, optical and sometimes radio) which follows the prompt emission, is called afterglow. The shape of a lightcurve (brightness in unit of time and detector area) of both prompt and afterglow of a typical GRB are shown in right panel of Figure 2.

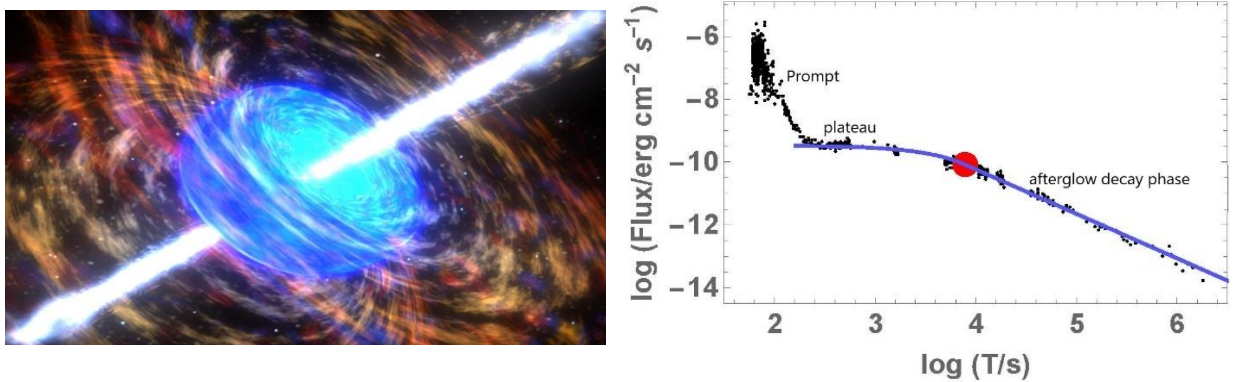

Fig. 2. Left panel: The explosion of a massive star producing a GRB. Credit to NASA. Right panel:A GRB lightcurve (brightness in unit of time and detector area) as a function of time. This picture shows the energy range of
The blue line represents the Willingale et al. 2007 model for the plateau emission. The red dot identifies the end of the plateau emission

Notwithstanding the variety of GRB peculiarities, some common features may be identified by looking at their light curves. A crucial breakthrough in this area has been the observation of GRBs by the Neil Gehrels Swift Observatory (Swift; Gehrels et al., 2004), launched in 2004. With its instruments on board, Swift provides a rapid follow-up of the afterglows in several wavelengths (hard and soft X-rays and optical) with better coverage than previous missions. Swift observations have revealed a more complex behavior of the lightcurves afterglow (O'Brien et al. 2006, Sakamoto et al. 2007) than in the past. They can be divided into two, three and even more segments in the afterglows. The second segment, when it is flat, is called plateau emission (see Figure 2).

The plateau observed in X-rays generally lasts from $10^2$ to $10^5$ s and is followed by a steeper power law decay phase. 42% of X-ray detections present plateau emission confirming Evans et al. (2009). The favored model for the lightcurves is a relativistic blast wave plugging into the circumstellar medium (see, e.g., MacFadyen et al. 2001) and thus producing an external shock (ES).

A general problem is to test several theoretical scenarios and pinpoint which are the most plausible models that can explain the majority of GRB observations. However, we should try to understand to which extant some models are preferred in the study and in the analysis based on our prior beliefs.

Discovering universal relations linking prompt and afterglow properties is a crucial step for the theoretical interpretation of the GRB physics. A general problem is to explore extensively data to explain the afterglow, and prompt-afterglow correlations observed in several wavelengths, from high-energy γ-rays to the optical band. The use of correlations enables the astrophysics community to understand GRBs as unique phenomena by relying on statistical samples rather than on individual GRBs, which are often complex in behavior and may not be representative of the GRB population. For such an analysis it is important to include all GRB classes, but to treat each class separately because, as we have pointed out, different physical scenarios may characterize each sub-class. Such kind of analysis will pinpoint through correlations any commonalities between GRBs and their subclasses, and will assist in their interpretation, with the aim of uncovering the most plausible explanation for the GRB emission mechanism. In addition, the final aim is to use one or more GRB classes as cosmological standard candles (Dainotti et al. 2015b). Standard candles are astronomical objects whose luminosity is known or can be derived from other distance-independent observables. Thus, casting light on the emission mechanism of GRBs will highly likely pinpoint a class of objects which can be ascribed to the same theoretical scenario and thus can be used as standard set for using GRBs as cosmological tools. Although many GRB correlations have been studied extensively, three crucial points are missing: they have not been investigated in a multiwavelength domain, observations have not been catalogued against several theoretical models in a statistical way (only a few peculiar GRBs have been extensively tested within several scenarios) and they have not been corrected for instrumental selection biases and cosmological evolution (evolution of the GRB properties with redshift). One of the analysis that has been done in a statistical direction is to test within the magnetar model (Stratta, Dainotti et al. 2018) a subset of 40 GRBs with relatively flat plateaus like the ones shown in Figure 2. Therefore, it is imperative to first determine the true correlations among the variables, not those introduced by the observational selection effects. Once these issues will be fully addressed, these correlations become reliable model discriminators and cosmological tools.

## 7. Ontological categorizations and the case of Gamma-Ray Bursts

One of the efforts the Gamma-Ray Burst community nowadays is to pinpoint a specific class or categories of GRBs with peculiar properties that can be used as standard candles, astronomical objects whose luminosity is known or can be derived through phenomenological correlations among relevant intrinsic properties.

These classes of GRBs are made possible under the assumption of a categorial framework or a conceptual scheme. More specifically, some GRBs can be chosen to belong to some categorial framework namely some classes of objects such as X-ray Flashes, GRBs associated with SNe, long GRBs, short GRBs and short GRBs with extended emission are categorized as different due to their phenomenological properties which turns to be related to their progenitor emission mechanism. These categorial schemes allows us to reconstruct retrospectively the conceptual framework that ground them. In addition, besides the phenomenological classification we have embarked in morphological categorization which can be ascribed to a particular categorial framework. In details, if we have a set of entities, $I_j$, namely GRBs, and a set of specific properties which can be pertinent to some of the GRBs, $P_i$, we need to tackle several issues in the classification itself.

If we try to reconstruct the links between entities and kinds, recalling the notion of properties sharing, we need to face the fact that some GRBs are contemporaneously X-Ray Flashes, ultra-long and associated with SNe. One of the challenges is how to disentangle between common properties shared by different classes and the conceptual classification pertinent to each entity. Namely, if we have a given GRB which is X-Ray Flashes, ultra-long and associated with SNe what will be the progenitor mechanism that can be plausible for that kind of entity? What are the kinds of entities that can be determined through this classification?

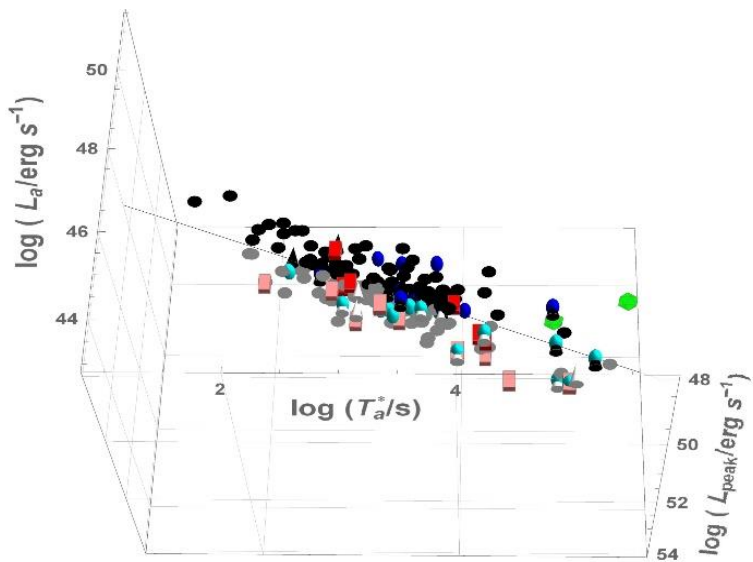

Figure 3
*The fundamental plane relation with several GRB classes: GRB-SNe (white cones), XRFs (blue spheres), Short with extended emission (red cuboid), Long GRBs (black circles) and Ultra Long (green rhomboids). Figure takem from Dainotti et al. 2017b*

Willingale at al. (2007) showed that Swift prompt and afterglow light curves may be fitted by the same analytical expression shown with a blue curve in figure 2. This provides the unprecedented opportunity to look for universal features that would allow us to recognize if GRBs are standard candles. Within this framework Dainotti et al. (2008) discovered a correlation for long GRBs between the X-ray luminosity at the end of the plateau phase, $L_X$ and its duration, $T^*_a$ (recently named in the literature as the Dainotti relation and hereafter called the $L_X$-$T^*_X$ correlation), later updated within the following works: Dainotti et al. 2010, 2011a, 2015a, 2017a. The $L_X$-$T^*_X$ relation implies that the more luminous the plateau phase is, the shorter its duration, i.e., the more powerful the phenomenon is, the quicker it consumes its energy. Being the slope of the correlation close to -1, thus the energy reservoir of the plateau emission is constant. This correlation has been a valuable tool to discriminate between the accretion onto a BH (e.g. Kumar et al. 2008a) and a millisecond spinning NS or a magnetar (e.g. Dall' Osso et al. 2011, Rowlinson et al. 2014, Rea et al. 2015). These models have been tested against this correlation and they are be able to reproduce it fairly well, but with scatter that pertains both error measurements and the intrinsic parameters of the given model. For example, the $L_{peak}$-$L_X$ relation can be reproduced within the standard ES model (Hascoet et al. 2014). The GRB fundamental plane relation could be ascribed to the magnetar scenario (Stratta et al. 2018) or the up-dust scattering model (Shao et al. 2009).

This correlation has been used as a cosmological probe (Cardone et al. (2009, 2010), Postnikov et al. 2014, Dainotti et al. 2013b). Moreover, the analysis performed by studying other physical parameters for both the prompt and afterglow phases, have led to the discovery of new significant correlations between the afterglow luminosity, $L_X$, and the peak luminosities of the prompt emission, $L_{peak}$ (Dainotti et al. 2011b, Dainotti et al. 2015b).

A combination of these two bi-variate relations, $L_{peak}$-$L_X$ and $L_X$-$T^*_X$, has led to a new physical tri-variate relation, $L_{peak}$-$L_X$-$T^*_X$ (Dainotti et al. 2016, 2017b), which defines a GRB fundamental plane, see Fig. 3.

For reviews on GRB correlations see Dainotti & del Vecchio 2017, Dainotti et al. 2018, Dainotti & Amati 2018, Dainotti 2019.

The fundamental plane is the tightest among the prompt-afterglow correlations and thus an excellent candidate to be used as a cosmological tool and as a theoretical model discriminator.

The interpretation of these correlations regarding their physical origins and their relationship to each other is still under debate, but several models have been proposed. We need to investigate whether the parameters that pertain particular models that can explain the proposed correlations are compatible with each other. Indeed, it is auspicious to highlight a coherent picture that explains all these correlations with a unique model.

### 8. Conclusions

The choices that we consider in the classification schemes may be driven by a relevant selection effect.

An example of selection biases is how we select the samples with given properties, as in extragalactic astronomy we have given sources with peculiar brightness and we have an all missing population of stars that are not detected due to the Malmquist bias effect, namely we see only the brightest sources at large distances, or for example on how catalogues are arranged. Why are we choosing a catalog rather than another? Are we driven by some beliefs? And is it possible to make these beliefs explicit? Can we quantify these beliefs in statistical terms, namely can we add an extra-variance that account specifically and exclusively on these beliefs?

If we manage to model this extra-variance we can have a control on how much results can change based on different assumptions.

Currently, GRB community is trying to properly disentangle among theoretical models trying to avoid the predilection of one model compared to another when both can explain the same phenomenology. How much the degree of degeneracy of the theoretical parameters is also influenced by the proposers?

As shown by the examples of SNe, galaxies and GRBs, just discussed, the cross-fertilization among astrophysical studies and philosophical conceptual analysis of ontological categorization can enlighten the relation between limits, selection and distortion in our understanding of empirical data, on one side, and the philosophical and metaphysical framework that is presupposed to make sense of our experience, on the other side. A careful and unconstrained analysis of ontological categorizations and metaphysical assumptions of our scientific enterprise will give us a better awareness of the fundamental theoretical choices in our relationship with experience. It will also advance the comprehension of how the metaphysical background theory grounds and drives our understanding of reality, questioning how the very notion of reality is affected by selection effect. The case of astrophysics and cosmology are paradigmatic examples, for they seem to be the most comprehensive, wide raging, exceptionally advanced fields and, still, connected to fundamental problems that go beyond scientific research to single out extremely general and essential questions.

**Acknowledgements**

M.G.Dainotti is grateful to the funds received by the American Astronomical Society Chretienne Fellowship and by MINIATURA2 Number 2018/02/X/ST9/03673. M.G. Dainotti & P. Valore are particularly grateful to Paola Sartorio, President of Fulbright Italia for favoring our meeting and connection and to the Fulbright Italia Alumni event in 2019.